\documentclass[vecphys]{svmult}

\usepackage{graphicx}
\graphicspath{{figures/}}

\DeclareGraphicsExtensions{.pdf}

\usepackage{multicol}        
\usepackage[bottom]{footmisc}
\usepackage{color}
\definecolor{grey}{rgb}{0.95, 0.95, 0.95}
\usepackage[final]{listings}
\usepackage{array}
\usepackage[font=footnotesize]{subfig}
\makeindex 

\begin{document}

\title*{LIKWID: Lightweight Performance Tools}
\titlerunning{Introduction and case studies}
\author{Jan Treibig \and Georg Hager \and Gerhard Wellein}
\authorrunning{J. Treibig et al.}
\institute{
J. Treibig $\cdot$ G. Hager $\cdot$ G. Wellein
\at Erlangen Regional Computing Center (RRZE), 
Friedrich-Alexander Universit\"at Erlangen-N\"urnberg, 
Martensstr. 1, D-91058 Erlangen, Germany\\
\email\texttt{\{jan.treibig,georg.hager,gerhard.wellein\}@rrze.uni-erlangen.de}
}
\maketitle

\abstract{
    Exploiting the performance of today's microprocessors requires intimate
    knowledge of the microarchitecture as well as an awareness of the
    ever-growing complexity in thread and cache topology. LIKWID is a set of
    command line utilities that addresses four key problems: Probing the thread
    and cache topology of a shared-memory node, enforcing thread-core affinity
    on a program, measuring performance counter metrics, and microbenchmarking
    for reliable upper performance bounds. Moreover, it includes an mpirun
    wrapper allowing for portable thread-core affinity in MPI and hybrid MPI/threaded
    applications. To demonstrate the capabilities of the tool set we show the
    influence of thread affinity on performance using the well-known OpenMP
    STREAM triad benchmark,  use hardware counter tools to study the
    performance of a stencil code, and  finally show how to detect bandwidth problems
    on ccNUMA-based compute nodes.
} 

\section{Introduction}
\label{sec:intro}

\begin{figure}[htbp]\centering
    \subfloat[Core 2 Quad]{\includegraphics*[width=0.4\linewidth]{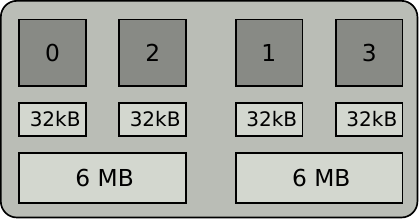}}\qquad
    \subfloat[Nehalem EP Westmere]{\includegraphics*[width=0.5\linewidth]{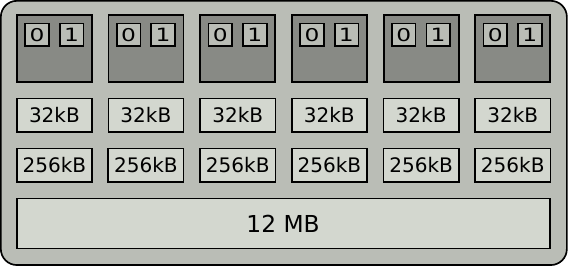}}
    \caption{Cache and thread topology of Intel Core 2 Quad and Nehalem EP Westmere processors.}
    \label{fig:topology}
\end{figure}

Today's multicore x86 processors bear multiple complexities when aiming for
high performance. Conventional performance tuning tools like Intel VTune,
OProfile, CodeAnalyst, OpenSpeedshop, etc., require a lot of experience in
order to get sensible results. For this reason they are usually unsuitable for
the scientific users, who would often be satisfied with a rough overview of the
performance properties of their application code. Moreover, advanced tools
often require kernel patches and additional software components, which make
them unwieldy and bug-prone. Additional confusion arises with the complex
multicore, multicache, multisocket structure of modern systems (see
Fig.~\ref{fig:topology}); users are all too often at a loss about how
hardware thread IDs are assigned to resources like cores, caches, sockets and
NUMA domains. Moreover, the technical details of how threads and processes are
bound to those resources vary strongly across compilers and MPI libraries.

LIKWID (``Like I Knew What I'm Doing'') is a set of easy to use command line
tools to support optimization. It is targeted towards performance-oriented
programming in a Linux environment, does not require any kernel patching, and
is suitable for Intel and AMD processor architectures. Multithreaded and even
hybrid shared/distributed-memory parallel code is supported. LIKWID comprises
the following tools:
\begin{itemize}
\item \verb.likwid-features. can display and alter the state of the
  on-chip hardware prefetching units in Intel x86 processors. 
\item \verb.likwid-topology. probes the hardware thread and cache
  topology in multicore, multisocket nodes. Knowledge like this is
  required to optimize resource usage like, e.g., shared caches and
  data paths, physical cores, and ccNUMA locality domains in parallel
  code.
\item \verb.likwid-perfCtr. measures performance counter metrics over
  the complete runtime of an application or, with support from a simple API,
  between arbitrary points in the code. Although it is possible to specify the
  full, hardware-dependent event names, some predefined event sets simplify
  matters when standard information like memory bandwidth or Flop counts is
  needed.
\item \verb.likwid-pin. enforces thread-core affinity
  in a multi-threaded application ``from the outside,'' i.e., without
  changing the source code. It works with all threading models that
  are based on POSIX threads, and is also compatible with hybrid 
  ``MPI+threads'' programming. Sensible use of likwid-pin requires
  correct information about thread numbering and cache topology, 
  which can be delivered by likwid-topology (see above).
\item \verb.likwid-mpirun. allows to pin a pure MPI or hybrid MPI/threaded
    application to dedicated compute resources in an intuitive and portable way.
\item \verb.likwid-bench. is a microbenchmarking framework allowing rapid
    prototyping of small assembly kernels. It supports threading, thread
    and memory placement, and performance measurement. 
    \verb.likwid-bench. comes with a wide range of
    typical benchmark cases and can be used as a stand-alone benchmarking
    application.
\end{itemize}
Although the six tools may appear to be partly unrelated, they solve the
typical problems application programmers encounter when porting and running
their code on complex multicore/multisocket environments.
Hence, we consider it a natural idea to provide them as a single
tool set.

This paper is organized as follows. Section~\ref{sec:tools} describes two of 
the tools in some detail and gives hints for typical use.
Section~\ref{sec:cases} demonstrates the use of LIKWID in three
different case studies, and Section~\ref{sec:conc} gives a summary and
an outlook to future work.

\section{Tools}
\label{sec:tools}

LIKWID only supports x86-based processors. Given the strong prevalence
of those architectures in the HPC market (e.g., 90\% of all systems in
the latest Top 500 list are of x86 type) we do not consider this a severe
limitation. In other areas like, e.g., workstations or desktops,
the x86 dominance is even larger. 

An important concept shared by all tools in the set is logical numbering
of compute resources inside so-called \emph{thread domains}. Under the Linux OS,
hardware threads in a compute node are numbered according to some scheme
that heavily depends on the BIOS and kernel version, and which may be unrelated
to natural topological units like cache groups, sockets, etc. 
Since users naturally think in terms of topological structures, LIKWID introduces 
a simple and yet flexible syntax for
specifying processor resources. This syntax consists of a prefix character and a list
of logical IDs, which can also include ranges.
The following domains are supported:
\begin{center}
\begin{tabular}{l@{\hspace*{5mm}}l}
Node & \verb+N+ \\
Socket & \verb+S[0-9]+  \\
Last level shared cache & \verb+C[0-9]+ \\
NUMA domain & \verb+M[0-9]+ 
\end{tabular}
\end{center}
Multiple ID lists can be combined, allowing a flexible numbering of compute resources.
To indicate, e.g., the first two cores of NUMA domains 1 and 3, the following 
string can be used:
\verb+M0:0,1@M2:0,1+.

In the following we describe two of the six tools in more detail. A thorough documentation
of all tools apart from the man pages is found on the WIKI pages on the LIKWID homepage \cite{likwid}.

\subsection{likwid-perfctr}

Hardware-specific optimization requires an intimate knowledge of the
microarchitecture of a processor and the characteristics of the code. While
many problems can be solved with profiling, common sense, and runtime measurements,
additional information is often useful to get a complete picture. 

Performance counters are facilities to count hardware events during code
execution on a processor. Since this mechanism is implemented directly in
hardware there is no overhead involved. All modern processors provide hardware
performance counters. They are attractive for application programmers, because
they allow an in-depth view on what happens on the processor while running
applications. As shown below,  \verb.likwid-perfctr. has practically
zero overhead since it reads performance metrics
at predefined points. It does not support statistical counter sampling.
At the time of writing, \verb.likwid-perfctr. runs on all current x86-based
architectures.

The probably best known and widespread existing tool is the PAPI library
\cite{PAPI,PAPI-C}.  A lot of research is targeted towards using hardware counter
data for automatic analysis and detecting potential performance
bottlenecks \cite{1213183,1418278,DBLP:conf/parco/GerndtFK05}. However, those
solutions are often too unwieldy for the common user, who would prefer a quick
overview as a first step in performance analysis. A key design goal for
\verb.likwid-perfctr. was ease of installation and use, minimal system requirements
(no additional kernel modules and patches), and --- at least for basic
functionality --- no changes to the user code.  A prototype for the development
of \verb.likwid-perfctr. is the SGI tool ``perfex,'' which was available on MIPS-based
IRIX machines as part of the ``SpeedShop'' performance suite. Cray provides a
similar, PAPI-based tool (craypat) on their systems~\cite{craypat}. 
\verb.likwid-perfctr. is a
dedicated command line tool for programmers, allowing quick and flexible
measurement of hardware performance counters on x86 processors, and is
available as open source.  It allows simultaneous measurements on multiple
cores.  Events that are shared among the cores of a socket (this pertains to
the ``uncore'' events on Core i7-type processors) are supported via ``socket
locks,'' which enforce that all uncore event counts are assigned to one thread
per socket.  Events are specified on the command line, and the number of events
to count concurrently is limited by the number of performance counters on the
CPU. These features are available without any changes in the user's source
code.  A small instrumentation (``marker'') API allows one to restrict
measurements to certain parts of the code (named regions) with automatic
accumulation over all regions of the same name. An important difference to most
existing performance tools is that event counts are strictly core-based instead
of process-based: Everything that runs and generates events on a core is taken
into account; no attempt is made to filter events according to the process that
caused them. The user is responsible for enforcing appropriate affinity to get
sensible results.  This could be achieved with \verb.likwid-perfctr. itself or
alternatively via \verb.likwid-pin. (see below for more information):
\begin{lstlisting}[basicstyle=\footnotesize\ttfamily]
$ likwid-perfctr -C S0:0 \
   -g SIMD_COMP_INST_RETIRED_PACKED_DOUBLE:PMC0,\
      SIMD_COMP_INST_RETIRED_SCALAR_DOUBLE:PMC1 ./a.out
\end{lstlisting}
(See below for typical output in a more elaborate setting.) In this example,
the computational double precision packed and scalar SSE  retired instruction
counts on an Intel Core 2 processor are assigned to performance counters 0 and
1 and measured on the first core (ID 0) of the first socket (domain S0) over the duration of
\verb!a.out!'s runtime.  As a side effect, it becomes possible to use
\verb.likwid-perfctr. as a monitoring tool for a complete shared-memory node, just by
specifying all cores for measurement and, e.g., ``\verb.sleep.'' as
the application:
\begin{lstlisting}[basicstyle=\footnotesize\ttfamily]
$ likwid-perfctr -c N:0-7 \
   -g SIMD_COMP_INST_RETIRED_PACKED_DOUBLE:PMC0,\
      SIMD_COMP_INST_RETIRED_SCALAR_DOUBLE:PMC1 \
         sleep 1
\end{lstlisting}

Apart from naming events as they are documented in the vendor's
manuals, it is also possible to use preconfigured \emph{event sets} (groups)
with derived metrics. This provides a simple abstraction layer in
cases where standard information like memory bandwidth, Flops per
second, etc., is sufficient:
\begin{lstlisting}[basicstyle=\footnotesize\ttfamily]
$ likwid-perfctr -C N:0-3 -g FLOPS_DP  ./a.out
\end{lstlisting}
The event groups are partly inspired from a technical report published by AMD
\cite{AMD}, and all supported groups can be obtained by using the \verb.-a.
command line switch.  We try to provide the same preconfigured event groups on all
supported architectures, as long as the native events support them. This allows
the beginner to concentrate on the useful information right away, without the
need to look up events in the manuals (similar to PAPI's high-level events). 
In the usage scenarios described so far there is no interference of
\verb.likwid-perfctr. while user code is being executed, i.e., the overhead is very
small (apart from the unavoidable API call overhead in marker mode).

The following example illustrates the use of the marker API in a serial
program with two named regions (``\verb.Main.'' and ``\verb.Accum.''):
\begin{lstlisting}
  #include <likwid.h>
  ...
  int coreID = likwid_processGetProcessorId();
  likwid_markerInit(numberOfThreads,numberOfRegions);
  int MainId  = likwid_markerRegisterRegion("Main");
  int AccumId = likwid_markerRegisterRegion("Accum");
  
  likwid_markerStartRegion(0, coreID);
  // measured code region "Main" here
  likwid_markerStopRegion(0, coreID, MainId);
  
  for (j = 0; j < N; j++) {
     likwid_markerStartRegion(0, coreID);
     // measured code region "Accum" here
     likwid_markerStopRegion(0, coreID, AccumId);
  }
  likwid_markerClose();
\end{lstlisting}
Event counts are automatically accumulated on multiple calls. Nesting or
partial overlap of code regions is not allowed. The API requires specification of a
thread ID (0 for one process only in the example) and the core ID of the
thread/process. The LIKWID API provides simple functions to determine the core
ID of processes or threads.  The following listing shows the shortened output of
\verb.likwid-perfctr. after measurement of the \verb.FLOPS_DP. event group on four
cores of an Intel Core 2 Quad processor in marker mode with two named regions
(``\verb.Init.'' and ``\verb.Benchmark.,'' respectively):
\begin{lstlisting}[basicstyle=\ttfamily\tiny]
$ likwid-perfCtr -c 0-3 -g FLOPS_DP -m ./a.out
-------------------------------------------------------------
CPU type:       Intel Core 2 45nm processor
CPU clock:      2.83 GHz
-------------------------------------------------------------
Measuring group FLOPS_DP
-------------------------------------------------------------
%Region: Init% 
+--------------------------------------+--------+--------+--------+--------+
|                Event                 | core 0 | core 1 | core 2 | core 3 |
+--------------------------------------+--------+--------+--------+--------+
|          INSTR_RETIRED_ANY           | 313742 | 376154 | 355430 | 341988 |
|        CPU_CLK_UNHALTED_CORE         | 217578 | 504187 | 477785 | 459276 |
. . .
+--------------------------------------+--------+--------+--------+--------+
+-------------+-------------+-------------+-------------+-------------+
|   Metric    |   core 0    |   core 1    |   core 2    |   core 3    |
+-------------+-------------+-------------+-------------+-------------+
| Runtime [s] | 7.67906e-05 | 0.000177945 | 0.000168626 | 0.000162094 |
|     CPI     |  0.693493   |   1.34037   |   1.34424   |   1.34296   |
| DP MFlops/s |  0.0130224  | 0.00561973  | 0.00593027  | 0.00616926  |
+-------------+-------------+-------------+-------------+-------------+
%Region: Benchmark% 
+-----------------------+-------------+-------------+-------------+-------------+
|           Event       |   core 0    |   core 1    |   core 2    |   core 3    |
+-----------------------+-------------+-------------+-------------+-------------+
| INSTR_RETIRED_ANY     | 1.88024e+07 | 1.85461e+07 | 1.84947e+07 | 1.84766e+07 |
. . .
|     CPI     |  1.52023  |  1.52252   |  1.52708   |  1.52661   |
| DP MFlops/s |  1624.08  |  1644.03   |  1643.68   |   1645.8   |
+-------------+-----------+------------+------------+------------+
\end{lstlisting}
Note that the \verb.INSTR_RETIRED_ANY. and \verb.CPU_CLK_UNHALTED_CORE. events
are always counted (using two nonassignable ``fixed counters'' on the Core 2
architecture), so that the derived \verb.CPI. metric (``cycles per instruction'')
is easily obtained.

\subsection{likwid-pin}

\begin{figure}[tb]\centering
    \includegraphics*[width=0.6\linewidth]{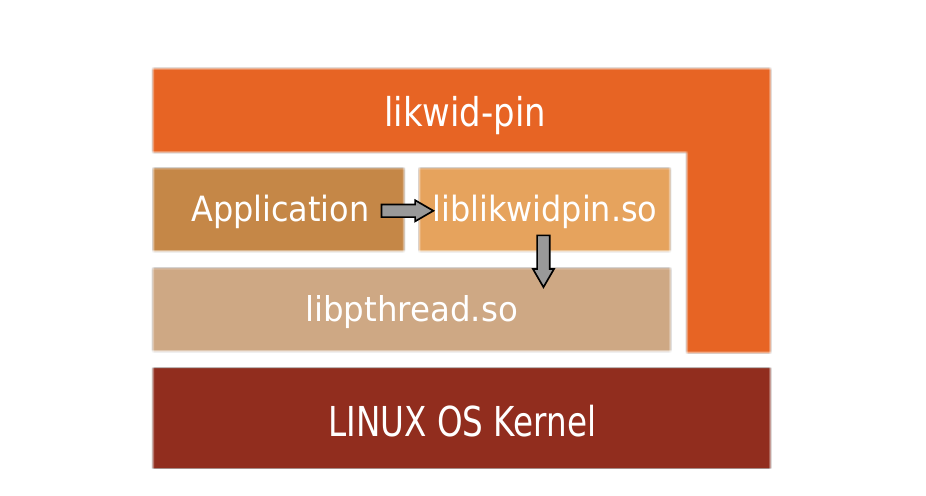}\hfill
\begin{minipage}[b]{0.4\linewidth}\caption{Basic architecture 
	of likwid-pin.}\label{fig:likwid-pin}\end{minipage}
\end{figure}
Thread/process affinity is vital for performance. If topology information is
available, it is possible to ``pin'' threads according to the application's resource
requirements like bandwidth, cache sizes, etc. Correct pinning is even more
important on processors supporting SMT, where multiple hardware threads share resources
on a single core. \verb.likwid-pin. supports thread affinity for all threading models that
are based on POSIX threads, which includes most OpenMP implementations. By
overloading the \verb+pthread_create+ API call with a shared library wrapper, 
each thread can be pinned in turn upon creation, working through a list of 
core IDs. This list, and possibly other parameters, are encoded in environment
variables that are evaluated when the library wrapper is first called. 
\verb.likwid-pin. simply starts the user application with the library
preloaded.

This architecture is illustrated in Fig.~\ref{fig:likwid-pin}. No code
changes are required, but the application must be dynamically linked. This
mechanism is independent of the processor architecture, but the way the compiled
code creates application threads must be taken into account: For instance, the
Intel OpenMP implementation  always runs \verb-OMP_NUM_THREADS- threads but
uses the first newly created thread as a management thread, which should not be
pinned. This knowledge must be communicated to the wrapper library. The following
example shows how to use \verb.likwid-pin. with an OpenMP application compiled with
the Intel compiler:
\begin{lstlisting}
$ export OMP_NUM_THREADS=4
$ likwid-pin -c N:0-3 %-t intel% ./a.out
\end{lstlisting}
In general, \verb.likwid-pin. can be used as a replacement for the \verb.taskset.
tool, which cannot pin threads individually.
Currently, POSIX threads, Intel OpenMP, and GNU (gcc) OpenMP are
supported directly, and the latter is assumed as the default if the
\verb.-t. option is not used. A bit mask can be specified to identify
the management threads for cases not covered by the available parameters
to the \verb.-t. option. Moreover, \verb.likwid-pin. can also be employed
for hybrid programs that combine MPI with some threading model, if 
the MPI process startup mechanism establishes a Linux cpuset for every
process.


The big advantage of \verb.likwid-pin. is its portable approach to the pinning
problem, since the same tool can be used for most applications, compilers, MPI
implementations, and processor types. In Section~\ref{sec:case_1} the usage
model is analyzed in more detail on the example of the STREAM triad.

\section{Case studies}
\label{sec:cases}

\subsection{Case study 1: Influence of thread topology on STREAM triad performance}
\label{sec:case_1}

%
\begin{figure}[btp]\centering
    \subfloat[not pinned]{\includegraphics*[width=0.49\linewidth]{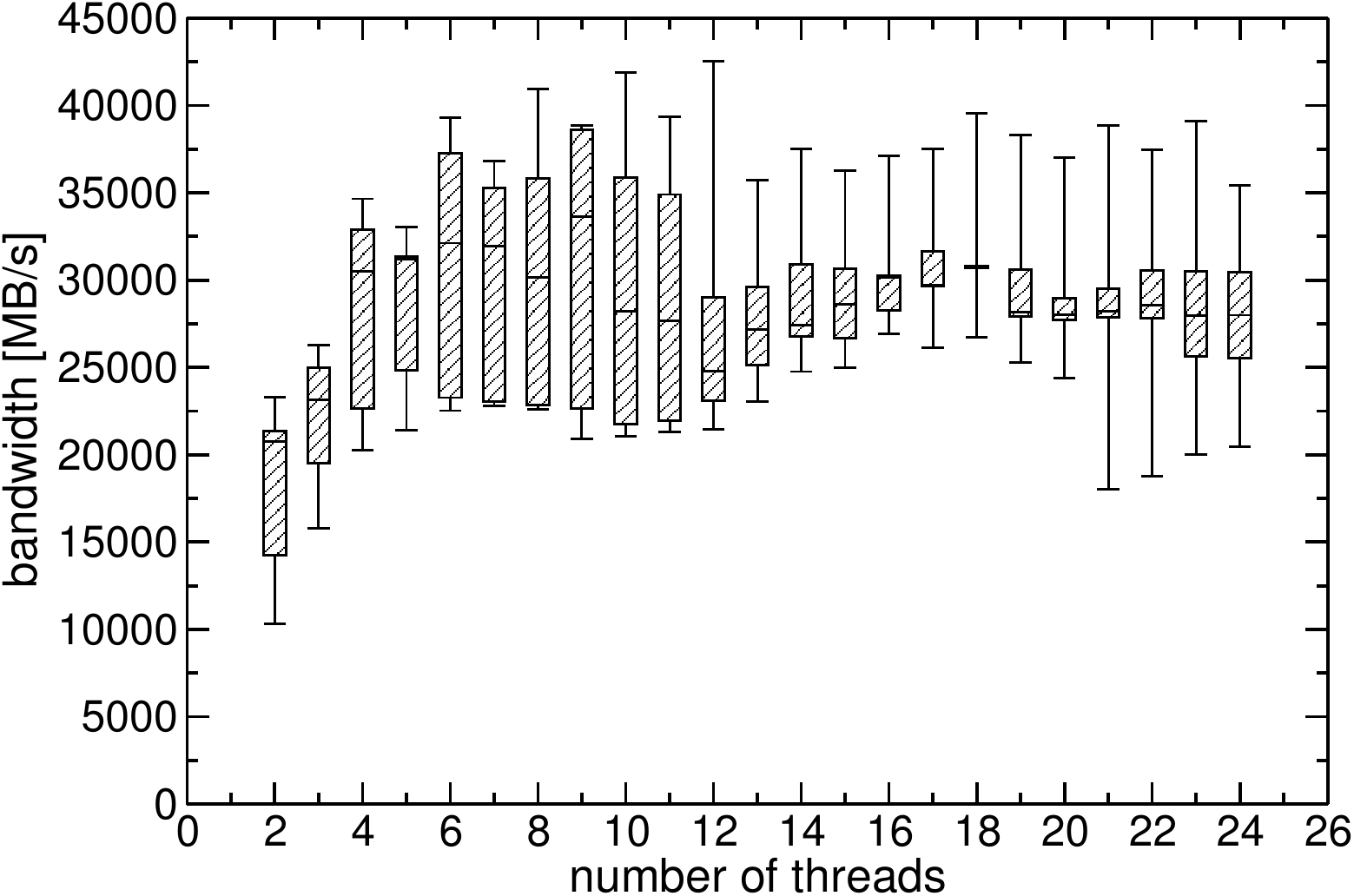}}\hfill
    \subfloat[pinned]{\includegraphics*[width=0.49\linewidth]{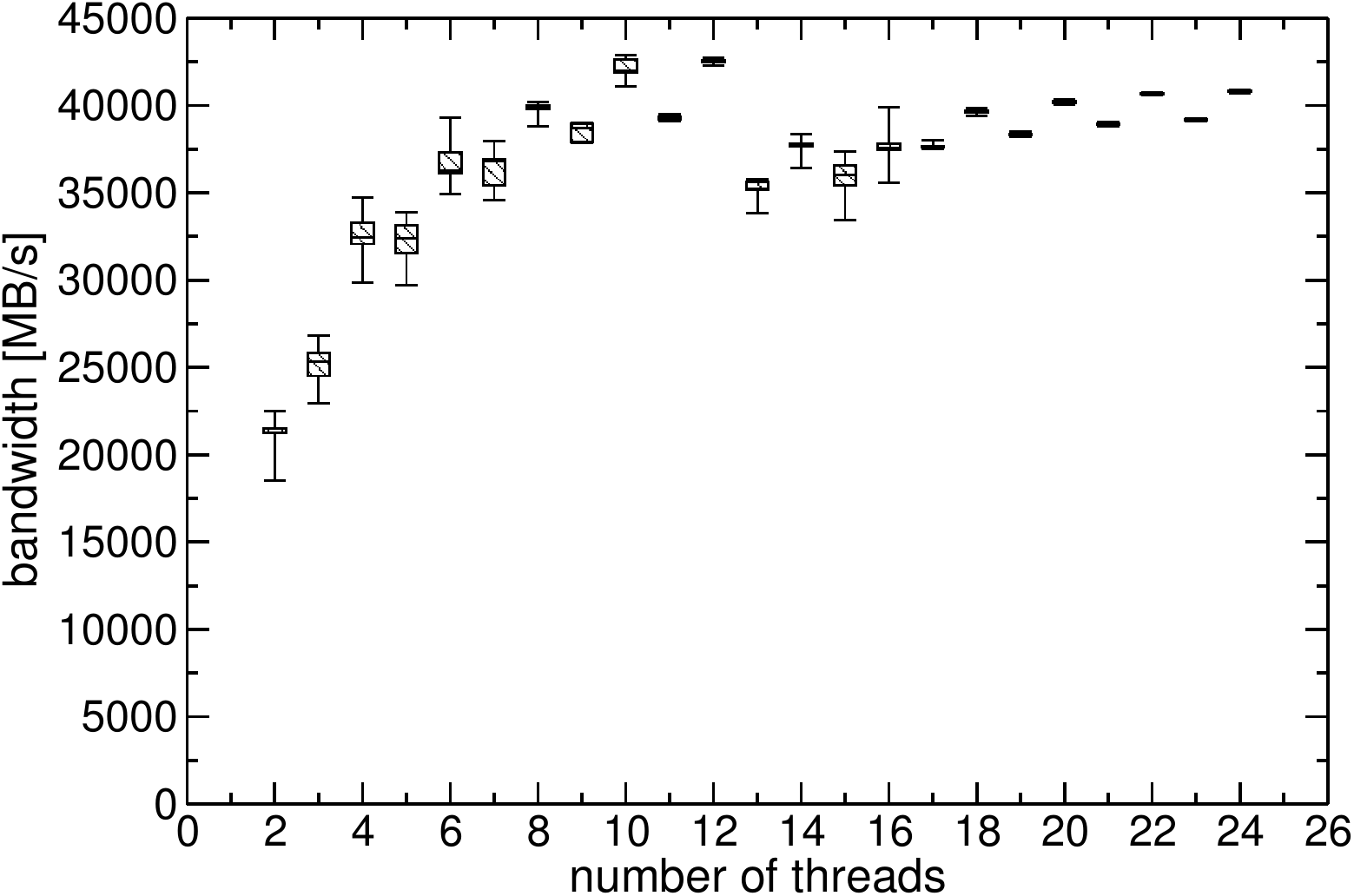}}
    \caption{STREAM triad test run with the Intel C compiler on a dual-socket
    Intel Westmere system (six physical cores per socket). In Fig. (a) threads
    are not pinned and the Intel pinning mechanism is disabled. In Fig. (b)
    the application is pinned such that threads are equally
    distributed on the sockets to utilize the memory bandwidth in the most
    effective way. Moreover, the threads are first distributed over physical
    cores and then over SMT threads.}
    \label{fig:stream_icc}
\end{figure}
%
%
To illustrate the general importance of thread affinity we use the well known
OpenMP STREAM triad on an Intel Westmere dual-socket system. Intel Westmere is
a hexacore design based on the Nehalem architecture and supports two SMT
threads per physical core. The Intel C compiler version 11.1 was used with options
\verb+-openmp+ \verb+-O3+ \verb+-xSSE4.2+ \verb+-fno-fnalias+.  Intel
compilers support thread affinity only if the application is executed on Intel
processors. The functionality of this topology interface is controlled by
setting the environment variable \verb+KMP_AFFINITY+. In our tests
\verb+KMP_AFFINITY+ was set to \verb+disabled+.  For the case of the STREAM
triad on these ccNUMA architectures the best performance is achieved if threads
are equally distributed across the two sockets.

Figure~\ref{fig:stream_icc} shows the results.  The non-pinned case shows a
large variance in performance especially for the smaller thread counts where
the probability is large that only one socket is used. With larger thread
counts there is a high probability that both sockets are used, still there is
also a chance that cores are oversubscribed, which reduces performance.
The pinned case consistently shows high performance throughout.
It is apparent that the SMT threads of Westmere increase the
chance of different threads fighting for common resources. 
\subsection{Case Study 2: Monitoring the performance of a Lattice Boltzmann fluid solver}
\label{sec:case_2}

\begin{figure}[btp]\centering
    \subfloat[Memory bandwidth]{\includegraphics*[width=0.49\linewidth]{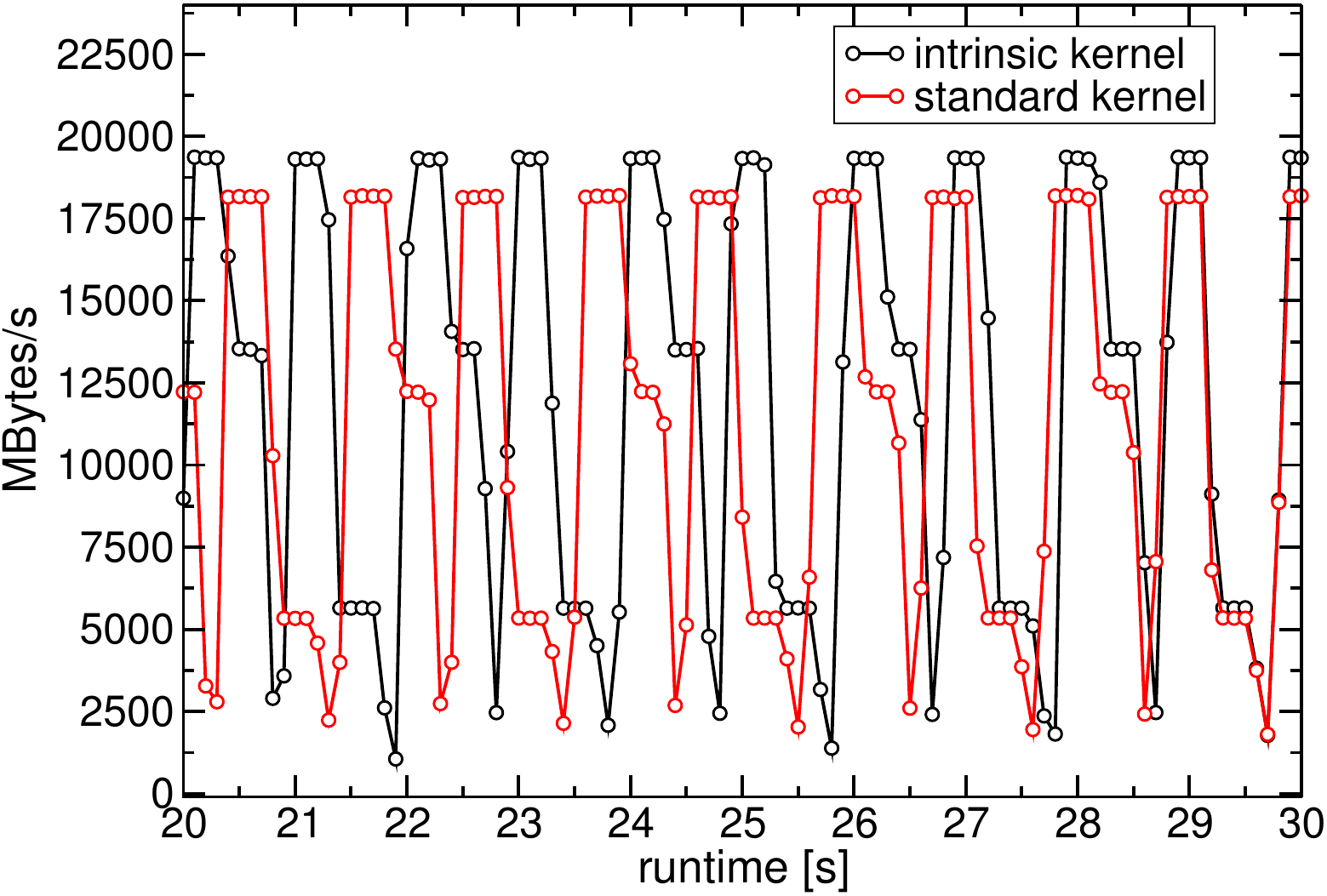}}\hfill
    \subfloat[Floating point performance]{\includegraphics*[width=0.49\linewidth]{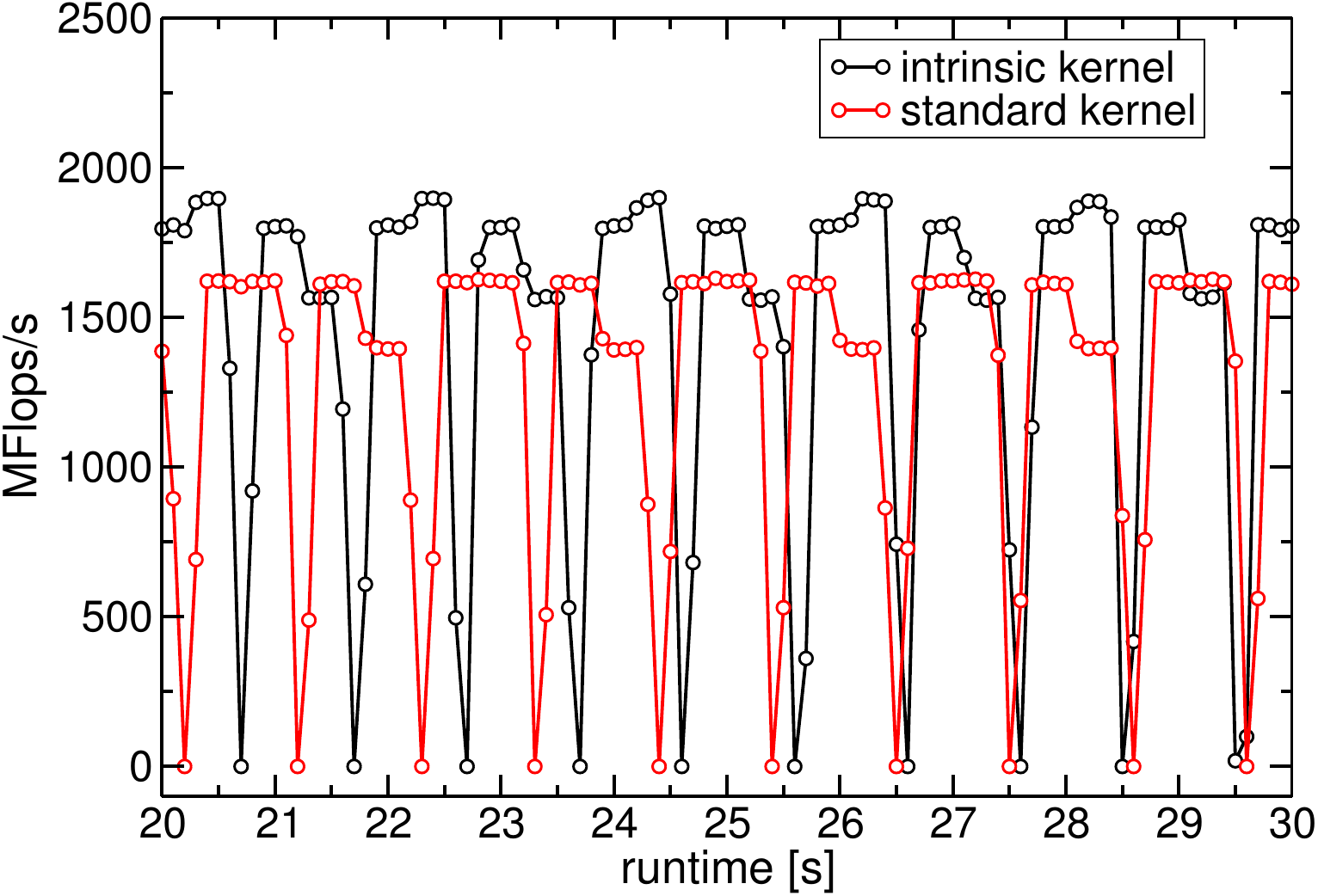}}
    \caption{Time-resolved results for the iteration phase of an MPI-parallel Lattice
    Boltzmann solver on one socket (four cores) of an Intel Nehalem compute node. 
	The compute performance in MFlops/s (Fig. (a)) and the memory bandwidth 
	in MBytes/s (Fig. (b)) are shown over a duration of 10 seconds, comparing
	two versions of the computational kernel (standard C versus SIMD
	intrinsics).}
    \label{fig:lbm_trace}
\end{figure}
To demonstrate the daemon mode option of \verb.likwid-perfctr. an
MPI-parallel Lattice Boltzmann
fluid solver was analyzed on a Intel Nehalem quad-core system (Fig.~\ref{fig:lbm_trace}). The daemon mode
of \verb.likwid-perfctr. allows time-resolved measurements of counter values and
derived metrics in performance groups. It is used as follows:
\begin{lstlisting}
$ likwid-perfctr -c S0:0-3 -g FLOPS_DP %-d 800ms%
\end{lstlisting}
This command measures the performance group \verb+FLOPS_DP+ on all physical
cores of the first socket, with an interval of 800\,ms between samples. 
\verb.likwid-perfctr. will only read out the hardware
monitoring counters and print the difference between the current and the
previous measurement.  Therefore, the overhead is kept
to a minimum. For this analysis the performance groups \verb+FLOPS_DP+ and
\verb+MEM+ were used.

\subsection{Case Study 3: Detecting ccNUMA problems on modern compute nodes}
\label{sec:case_3}

\begin{figure}[tbp]\centering
    \subfloat[Sequential initialization]{\includegraphics*[width=0.3\linewidth]{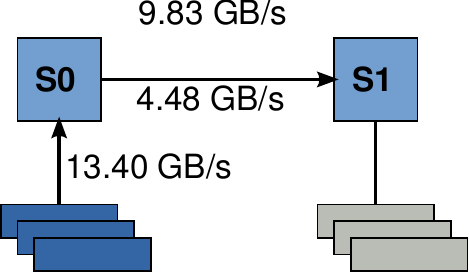}}\hfill
    \subfloat[First touch policy]{\includegraphics*[width=0.3\linewidth]{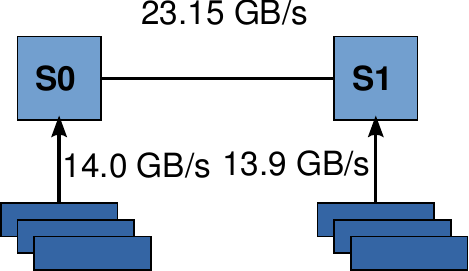}}\hfill
    \subfloat[Interleave policy]{\includegraphics*[width=0.3\linewidth]{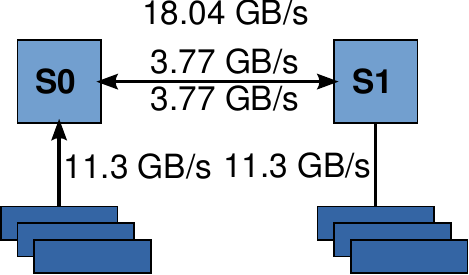}}
    \caption{NUMA problems reproduced with \texttt{likwid-bench} on the example of a
    memory copy benchmark on an Intel Nehalem dual-socket quad-core machine. The
    bandwidth on the top is the effective total application bandwidth as
    measured by \texttt{likwid-bench} itself. The other bandwidth values are from
    \texttt{likwid-perfctr} measurements in wrapper mode.}
    \label{fig:numa}
\end{figure}
Many performance problems in shared memory codes are caused by an
inefficient use of the ccNUMA memory organization on modern compute
nodes. CcNUMA technology achieves scalable memory size and bandwidth
at the price of higher programming complexity: The well-known locality
and contention problems can have a large impact on the performance of
multithreaded memory-bound programs if parallel first touch placement
is not used on initialization loops, or is not possible for some
reason~\cite{hpc4se}.


\verb.likwid-perfctr. supports the developer in detecting NUMA problems with two
performance groups: \verb.MEM. and \verb.NUMA.. While on some architectures like, 
e.g., newer Intel systems, all events can be measured in one run using the 
\verb.MEM. group, a separate group (\verb.NUMA.) is necessary on 
AMD processors.

The example in Fig.~\ref{fig:numa} shows results for a memory copy benchmark, 
which is part of
\verb.likwid-bench.. Since \verb.likwid-bench. allows easy control of thread and data
placement it is well suited to demonstrate the capabilities of \verb.likwid-perfctr.
in detecting NUMA problems. Here, \verb.likwid-perfctr. was used as follows:
\begin{lstlisting}
$ likwid-perfctr -c S0:0@S1:0 -g MEM ./a.out
\end{lstlisting}
The relevant output for the derived metrics could look like this:
\begin{lstlisting}
+-----------------------------+-----------+----------+
|           Metric            |  core 0   |  core 4  |
+-----------------------------+-----------+----------+
|         Runtime [s]         |  4.71567  | 0.138517 |
|             CPI             |  16.4815  | 0.605114 |
| Memory bandwidth [MBytes/s] |  6.9273   | 6998.71  |
|  Remote Read BW [MBytes/s]  | 0.454106  | 4589.46  |
| Remote Write BW [MBytes/s]  | 0.0705132 | 2289.32  |
|    Remote BW [MBytes/s]     | 0.524619  | 6878.78  |
+-----------------------------+-----------+----------+
\end{lstlisting}
All threads were executed on socket zero, as can be seen from the runtime which
is based on the \verb+CPU_CLK_UNHALTED_CORE+ metric. All program data
originated from socket one since there is practically 
no local memory bandwidth. Hence,
all bandwidth on socket one came from the remote socket.  Fig.~\ref{fig:numa}~(a) 
shows the results for sequential data initialization on one socket; the
overall bandwidth is 9.83\,GB/s. Fig.~\ref{fig:numa}~(b) shows the case with
correct first touch data placement on both sockets. The effective bandwidth is
23.15~GB/s, and the scalable ccNUMA system is used in the most efficient way.
If an application cannot be easily changed to make use of the 
first touch memory policy, a 
viable compromise is often to switch to automatic round-robin page
placement across a set of NUMA domains, or \emph{interleaving}.
\verb.likwid-pin. can enforce interleaving for all NUMA domains included in a
threaded run. This can be achieved with the \verb+-i+ option:
\begin{lstlisting}
$ likwid-pin -c S0:0-3@S1:0-3 -t intel %-i% ./a.out
\end{lstlisting}
Figure~\ref{fig:numa}~(c) reveals that the memory bandwidth achieved with
interleaving policy, while not as good as with correct first touch, is still
much larger than the bandwidth of case (a) with all data in one NUMA domain.

\section{Conclusion and future plans}
\label{sec:conc}

LIKWID is a collection of command line applications supporting
performance-oriented software developers in their effort to utilize today's
multicore processors in an effective manner. LIKWID does not try to follow the
trend to provide yet another complex and sophisticated tooling environment,
which would be difficult to set up and would overwhelm the average user with
large amounts of data. Instead it tries to make the important functionality
accessible with as few obstacles as possible. The focus is put on simplicity
and low overhead. \verb.likwid-topology. and \verb.likwid-pin. 
enable the user to account for
the influence of thread and cache topology on performance and pin their
application to physical resources in all possible scenarios with one single
tool and no code changes.  The usage of \verb.likwid-perfctr. was
demonstrated on two examples.  LIKWID is open source and released under GPL2.  It can be
downloaded at \verb+http://code.google.com/p/likwid/+. 

Future plans include applying the philosophy of LIKWID to other areas like,
e.g., profiling (also on the assembly level). Emphasis will also be put on a
further improvement with regard to usability. It is also planned to port parts
of LIKWID to the Windows operating system. An ongoing effort is to add support
for present and upcoming architectures like, e.g., the
Intel Sandy Bridge microarchitecture.

\section*{Acknowledgment}
We are indebted to Intel Germany for providing test systems and early access
hardware for benchmarking. A special acknowledgment goes 
to Michael Meier, who had the basic
idea for \verb.likwid-pin., implemented the prototype, and provided many useful
thoughts in discussions. This work was supported by the Competence Network for
Scientific and Technical High Performance Computing in Bavaria (KONWIHR) under
the project ``OMI4papps.''

\end{document}